# Stable artificial solid electrolyte interfaces for lithium batteries

Lin Ma[1], Mun Sek Kim[2], Lynden A. Archer[2, *]

[1]Department of Materials Science and Engineering, Cornell University, Ithaca, New York 14853-5201, USA.

[2]School of Chemical and Biomolecular Engineering, Cornell University, Ithaca, New York, 14853-5201, USA.

[*]email: laa25@cornell.edu

**Abstract**

A rechargeable lithium metal battery (LMB), which uses metallic lithium at the anode, is among the most promising technologies for next generation electrochemical energy storage devices due to its high energy density, particularly when Li is paired with energetic conversion cathodes such as sulfur, oxygen/air, and carbon dioxide. Practical LMBs in any of these designs remain elusive due to multiple stubborn problems, including parasitic reactions of Li metal with liquid electrolytes, unstable/dendritic electrodeposition at the anode during cell recharge, and chemical reaction of dissolved cathode conversion products with the Li anode. The solid electrolyte interface (SEI) formed between lithium metal and liquid electrolytes plays a critical role in all of these processes. We report on the chemistry and interfacial properties of artificial SEI films created by in-situ reaction of a strong Lewis Acid $AlI_3$ additive, Li metal, and aprotic liquid electrolytes. We find that these SEI films impart exceptional interfacial stability to a Li metal anode. We further show that the improvements come from at least three processes: (i) in-situ formation of Li-Al alloy, (ii) formation of a LiI salt layer on Li, and (iii) creation of a stable polymer thin film on the lithium metal anode.

Rechargeable batteries able to reliably store large amounts of electrochemical energy are needed to meet increasing demands for long-lasting, portable electrical energy storage technology for electronic devices, electric vehicles, and autonomous robotics.[1,2] Lithium-ion batteries (LIBs) are currently the technology of choice for meeting these needs, however, with current LIBs reaching the theoretical capacity limits set by the chemistry of their cathode and anode materials, a new generation of rechargeable batteries is urgently needed. The lithium metal anode has been described as the "Holy Grail" of energy storage systems due to its extremely high theoretical specific capacity (3860 mAh/g), low gravimetric density (0.59g/cm$^3$), lowest negative redox potential vs. standard hydrogen electrode (-3.040V), [1,3] and the large variety of high-capacity unlithiated materials it enables as legitimate choices for the battery cathode. Thus, by replacing the carbonaceous host material used as the anode in an LIB with metallic lithium, rechargeable lithium metal batteries (LMBs) with impressive theoretical specific energies become possible.[4,5] Among these cathode materials, sulfur with a theoretical capacity of 1675 mAh/g has attracted sustained scientific interest for a variety of reasons, including its low cost, low toxicity, high natural abundance, and the fact that it undergoes spontaneous electrochemical reactions with lithium that do not require catalysts.[2,6,7]

Unfortunately, uncontrollable dendritic lithium growth and limited Coulombic efficiency during Li deposition/stripping inherent in all batteries that utilize metallic lithium as anode have prevented broader practical applications. The formation and subsequent growth of lithium dendrites induced by inhomogeneous distribution of current density on the lithium metal anode may pierce the polymer separator, resulting in short circuit and subsequent thermal runaway of the cell.[3,5,8] In addition, lithium metal is very reactive and over many cycles of charge and discharge will react with liquid electrolyte in contact with the metal to form fresh solid electrolyte interfaces (SEI), which ultimately consume the electrolyte causing low cycling efficiency as the internal resistance of the cell diverges.[9,10] The already complicated chemistry at the interface of a lithium metal anode and liquid electrolyte is made even more complex when lithium metal is paired with sulfur in a lithium-sulfur (Li-S) battery, which makes these batteries an important platform for fundamental studies of how each of the degradation processes interact to cause premature failure of LMBs.

In Li-S batteries the unique chemistry and transport behavior of soluble lithium polysulfides (LiPS) generated during the electrochemical process lead to multiple additional parasitic chemical pathways in the SEI that consume Li, deplete the active anode material, and may also cause the interfacial resistance at the Li metal anode to become more inhomogeneous, which promotes rough dendritic Li deposition during cell recharge.[2,11] The products created by reduction of sulfur by lithium have been studied extensively and are now thought to include $Li_2S_n$ species with n value raging from 1 to 8.[2,6,12] Whereas the low order $Li_2S_n$ (n ≤ 2) are insoluble in most aprotic liquid electrolyte solvents, high order lithium polysulfides $Li_2S_n$ (n > 2) can dissolve into the electrolyte, which will cause the low utilization of active materials and their parasitic reaction with Li anode.[12,13] Soluble lithium polysulfides (LiPS) diffusing throughout the separator could react with the Li anode to form the insoluble and insulating sulfides on the surface of the lithium anode, increasing the overpotential and lowering both the efficiency and rate capability.[14] To solve the LiPS dissolution problem, several studies have considered novel cathode designs, including use of nanostructured carbons as sorbents in the cathode to provide physical trapping for LiPS,[13,15,16] specially designed additives to sequester LiPS via chemical interaction,[11,17] and polymer coatings of the cathode to provide additional transport and kinetic barriers for LiPS dissolution.[18,19] In a departure from this approach, a recent study by Ma et al.[20] showed that carbon nanotubes grafted with covalently attached polyethyleneimine (PEI) chains take advantage of kinetic and thermodynamic processes to provide exceptionally high resistance to dissolution of LiPS in liquid electrolytes. Even in that case, however, the authors reported that over sufficiently long times, some amount of LiPS dissolves in the electrolyte. It means that the problem of LiPS dissolution in a Li-S cell cannot be solved through clever engineering of the cathode alone because the preferred electrolytes (eg. Dioxalane (DOL), 1,2-dimethoxyethane (DME), Tetraglyme) have sufficiently high solubility for LiPS that there will always exist a chemical potential gradient between the cathode an electrolyte at equilibrium, which favors dissolution and loss of LiPS to the electrolyte.

Normally, the three open problems with the Li-S cell — unstable, dendritic deposition of Li at the anode during recharge and dissolution, shuttling of LiPS formed at the cathode, and uncontrolled reaction of dissolved LiPS with the anode to form resistive insoluble passivating sulfide layers on the anode — are addressed independently with novel materials designs suitable for either electrode. A strategy that simultaneously sequesters sulfur in a Li-S battery cathode, which protects the anode from reaction with dissolved LiPS, and which eliminates dendritic deposition of Li at the nucleation step is a long sought after strategy for enabling Li-S cells able to live up to the potential of this battery technology. Herein, we report that introduction of the Lewis acid $AlI_3$ in the electrolyte engenders multiple synergistic processes that enable Li-S cells with unprecedented stability and cycling efficiency. Our interest in AlI$_3$ originates from two fundamental attributes of the material: i) The I$^-$ ion has long been understood to play a special role in adsorption phenomenon in electrochemistry. [21,22] I$^-$ belongs to the Class IB adsorbents, which have remarkable surface affinity and exhibit stronger interactions with electrodes than those known for class IA ions and for simple electrostatic interactions. It is believed that this feature of Class 1B adsorbents originates from donation of electrons from the adsorbing anions to available orbitals on the electrode surface. As a result, Class IB anions can be adsorbed on either positively charged or negatively charged electrode surfaces. Thus, dissociation of AlI$_3$ salt additives in an electrolyte is expected to result in an I$^-$ rich SEI layers at both electrodes. At a Li anode, the I$^-$ bonds with Li to form a conformal LiI salt layer localized at the electrode surface. Such a LiI coating has been shown by recent Joint Density Functional (JDFT) calculations to be as effective as LiF in lowering the activation barrier for Li+ transport across the electrolyte-electrode interface, [23,24] allowing it to conduct Li ions while at the same time preventing direct contact and reaction between lithium metal anode and electrolyte solvents. ii). AlI$_3$ can enhance Li cycling performance by in-situ formation a Li-Al metal alloy layer at the anode.[25] Such alloys have long been thought to provide among the most efficient barriers to Li dendrite formation, resulting in the enhancement of Li cyclability.[26-28] Additionally, we discovered that Al$^{3+}$, as a strong Lewis acid, is an efficient initiator for polymerization of DOL.[29] The polymerization reaction produces an ion-conducting film on the surface of lithium metal, which we believe stabilizes the lithium metal against side reactions with the electrolyte.

**Scheme 1** summarizes our understanding of the specific processes by which $AlI_3$ performs its remarkable functions in a Li-S cell. As a proof-of concept, we first fabricated the artificial protective film on Li metal by using an electrochemical approach. The protected lithium metal derived from this process was used in an additive-free (i.e. no $LiNO_3$) electrolyte to evaluate the effectiveness of the surface modification in overcoming the range of challenges with Li-S cells discussed in the introduction. These results are presented throughout the remainder of the paper. The protective SEI layer on Li metal anode used in the demonstration was formed by cycling a symmetric lithium metal cell in a DOL/DME-1M LiTFSI electrolyte with $AlI_3$ as an additive.

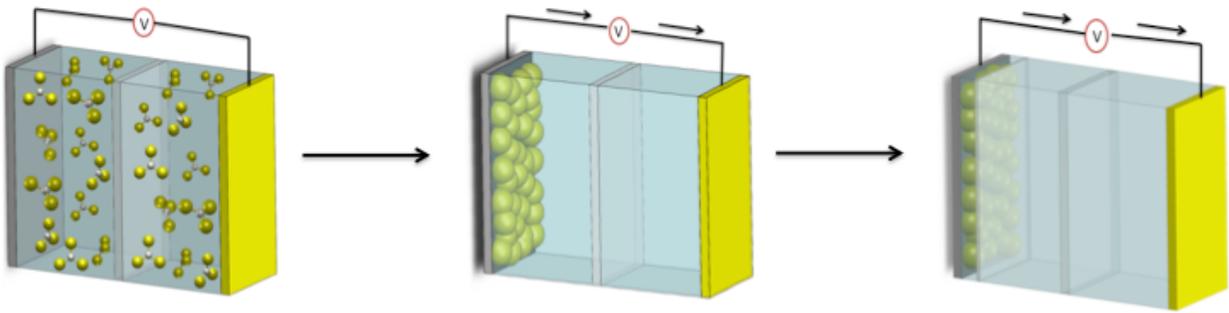

i. $3Li + AlI_3 \rightarrow 3LiI + Al$

ii. $Al + x\ Li^+ + xe^- \rightarrow Li_xAl$

iii.

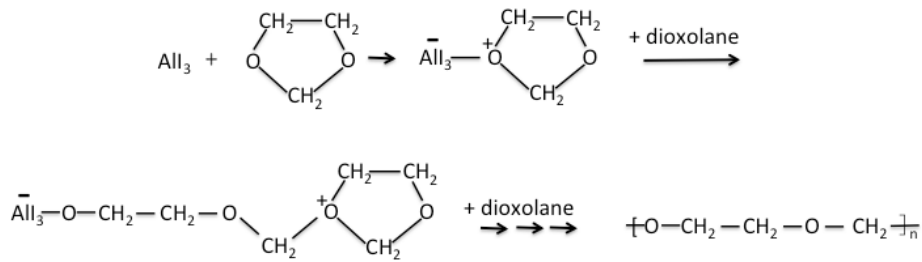

Scheme 1. Proposed formation mechanism for a multi-functional artificial solid electrolyte interfaces (SEI) at a Li anode produced by $AlI_3$.

In one approach, Li metal is first discharged, and then charged at constant current at 2% depth of discharge in the electrolyte containing $AlI_3$. This pretreatment allows the electrochemical cleaning of lithium metal and also the complete formation of stable SEI layer. The fast discharge we chose (2 mA/cm$^2$) can peel off the initial oxide film on the lithium surface, initializing the chemical reaction between pure Li and $Al^{3+}$.[30] The electrochemical process can also help to form Li-Al alloy on the lithium metal surface during the charge process. The voltage profile of the pretreatment step is provided in Supplemental Information as **Figure S1**. Photographic images of the lithium metal before and after treatment are shown in **Figure S2**.

**Results and discussion**

To evaluate the effectiveness of $AlI_3$ as a Li surface protection agent, lithium metal foil pretreated using the approach described above was used as electrodes in both symmetric (Li/Li) and full (Li-S) cells to investigate the effectiveness of the SEI layer in stabilizing the lithium metal against parasitic reactions with dissolved LiPS and dendrite formation on Li during cell recharge. Lithium metal is known to react spontaneously with LiPS. Once contacted, by an electrolyte containing dissolved LiPS species, lithium metal will be oxidized by LiPS to form solid $Li_2S$, which will deposit onto the lithium metal surface as an undesirable insulating layer. At the same time, the reduction of the dissolved LiPS leads to a distinct color change in the electrolyte as the order of the LiPS is reduced. To evaluate the effectiveness of our surface coating strategy, lithium foil with and without surface treatment was immersed in a solution of 0.05M $Li_2S_8$ dissolved in tetraglyme and the change of the electrolyte and lithium metal recorded. The results reported in the upper row of **Figure 1a** are for the pristine (untreated) Li foil, whereas those in the second row are typical results obtained using the $AlI_3$ treatment approach described above. For the pristine lithium, there is very obvious change of color of the electrolyte, indicating LiPS is reacting with lithium metal. The dark reddish color, which corresponds to high order LiPS, is observed to become markedly lighter over time. In contrast, the color of the electrolyte on the second row stays relatively dark, indicating LiPS is very stable in this case. After 12h, the electrolyte was taken out and characterized with UV-vis spectroscopy, and the results are shown in **Figure 1b**. The black curve corresponds to the initial $Li_2S_8$ solution in tetraglyme, which shows spectra consistent with literature results for basically long chain LiPS ($Li_2S_8$). [31,32] In the case of pristine lithium, the UV-vis signatures for short chain LiPS are clearly evident indicating the reduction of LiPS by reaction with lithium metal. And, consistent with the pictures shown in **Figure 1a**, the UV-vis spectrum of the electrolyte in contact with the pretreated Li foil is essentially identical to that of the freshly prepared electrolyte. The lithium metal immersed in the LiPS electrolyte for 12h was removed and characterized by X-ray diffraction (XRD); results are shown in **Figure 1c**. It is obvious that the pristine lithium reacts with LiPS and forms $Li_2S$ on the metal surface. However, for the pretreated lithium metal, XRD peaks for pure Li metal remain even after 12-hour exposure to LiPS and no obvious $Li_2S$ crystal structure is detected. Both the visualization of the color change of electrolyte, and the post-mortem analysis by UV-vis spectroscopy and XRD therefore confirm the stability of lithium metal against LiPS is improved substantially when it is pretreated with $AlI_3$.

The stability of the treated Li metal in the presence of LiPS was also investigated by Electrochemical Impedance Spectroscopy (EIS) experiments. Symmetric Li/Li cells either using pristine Li (control) or $AlI_3$ treated Li were employed in these experiments and the electrolyte is deliberately reinforced with 0.1M $Li_2S_8$ (**Figure 1d**), and the impedance is characterized as a function of time, which provides an indication of the lithium corrosion level by reacting with LiPS in the electrolyte. Both the real (resistance) and imaginary (capacitance) parts of the impedance are seen to increase more for the control cells. In particular the real part of the impedance of the control cells is seen to increase rapidly, reaching a value of around 250% of the initial value after 72 hours. In contrast, cells containing protected lithium are noticeably less reactive when in contact with LiPS rich electrolyte—only 45% increase is observed for the impedance over 72h. Following these measurements, the Li electrodes were harvested and the surface morphology of lithium metal observed using scanning electron microscopy (SEM). For the unprotected lithium metal, the lithium surface is seen to be very rough, indicating the severe erosion of lithium metal by LiPS; however the pretreated lithium has much less roughness (**Figure S3**).

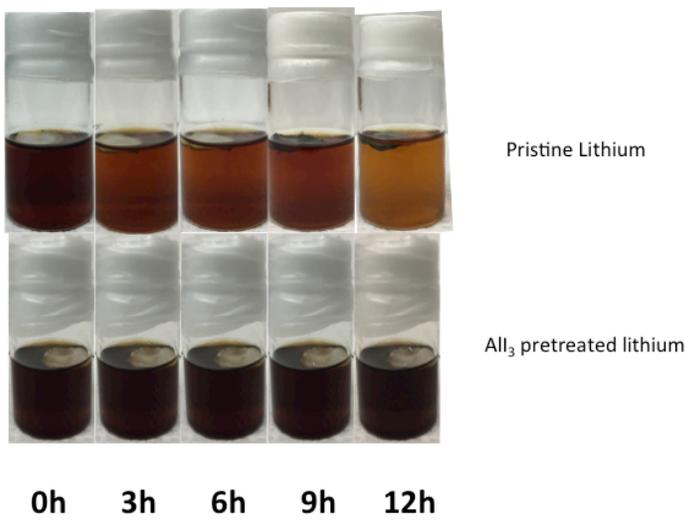

(a)

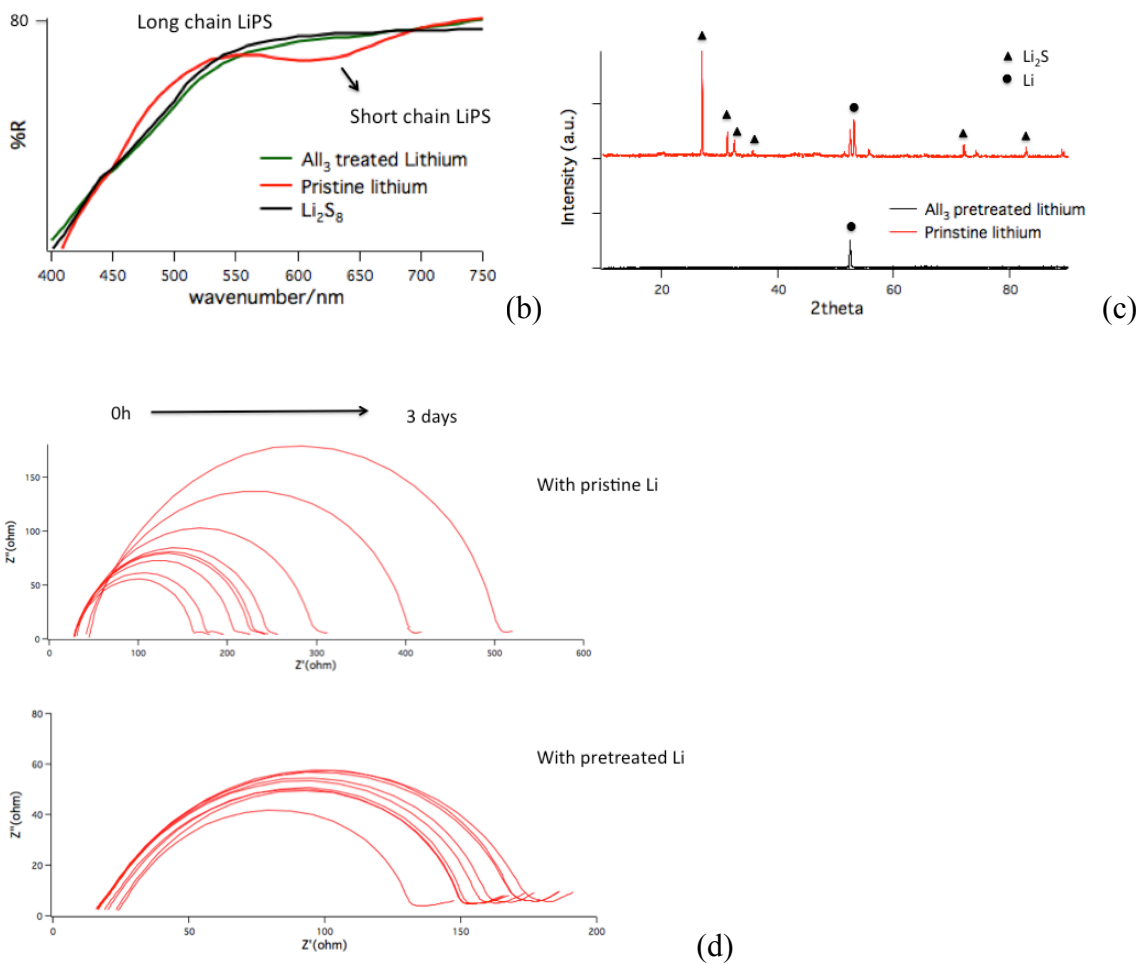

Figure 1. Stability of lithium metal against LiPS-rich electrolyte. a) Optical images recorded at a 3-hour interval of a LiPS-rich liquid electrolyte exposed to Li metal without (upper row) and with (lower row) pretreatment by $AlI_3$. b) Uv-vis spectra of a LiPS-rich electrolyte after exposure to lithium metal for 12 hours. c) XRD of the pristine lithium metal (red) and pretreated lithium metal foil (black) after immersion in a LiPS-rich electrolyte for 12 hours. d) Nyquist plot for a lithium symmetric cell containing a LiPS rich electrolyte recorded in 8 hour increments.

A remarkable and synergetic benefit of the $AlI_3$ surface treatment revealed by the proposed reaction mechanism in **Scheme 1** is that both the LiI layer and Li-Al alloys formed at the interface should stabilize Li metal against dendrite formation during cell cycling. To investigate these effects, symmetric (Li/Li) cells containing pristine and $AlI_3$- treated Li foil were assembled. A standard polypropylene membrane (Celgard™) was used as the separator and 1M LiTFSI in DOL/DME (v:v=1:1) was applied as electrolyte. To evaluate the stability of the cells to failure by dendrite-induced short circuiting, galvanostatic polarization measurements were performed in which lithium is continuously stripped from one electrode and plated on the other at a fixed current density; cell failure in this experiment occurs when the measured voltage is observed to drop discontinuously (see supporting **Figure S4**) as the internal short lowers the cell resistance.[33] **Figure 2a** reports the cell lifetimes at various current densities. It is seen that the, lifetime or short-circuit time $t_c$ is greatly improved when lithium metal is protected with SEI layer involving $AlI_3$. The improvement is more obvious when higher current is applied and the resultant $t_{sc}$ values at 3mA/cm$^2$ are the highest reported in the literature for this experiment.[3,33] It should also be noted that the electrolyte used in these experiments contains no additives, which means that the SEI layer is quite stable by itself over lithiation over a long period.

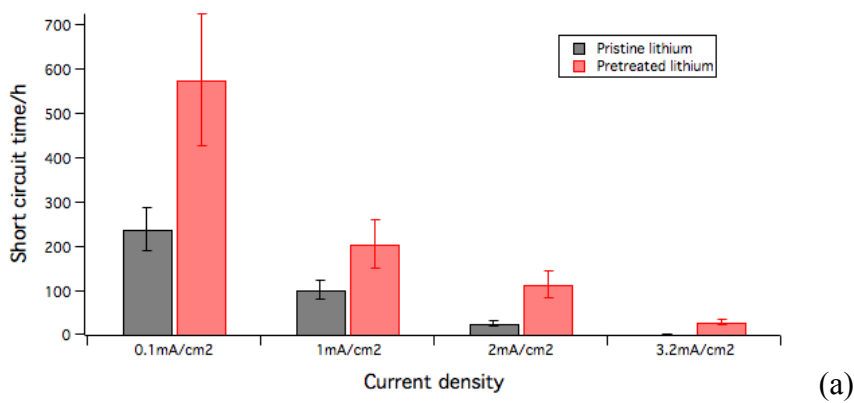

(a)

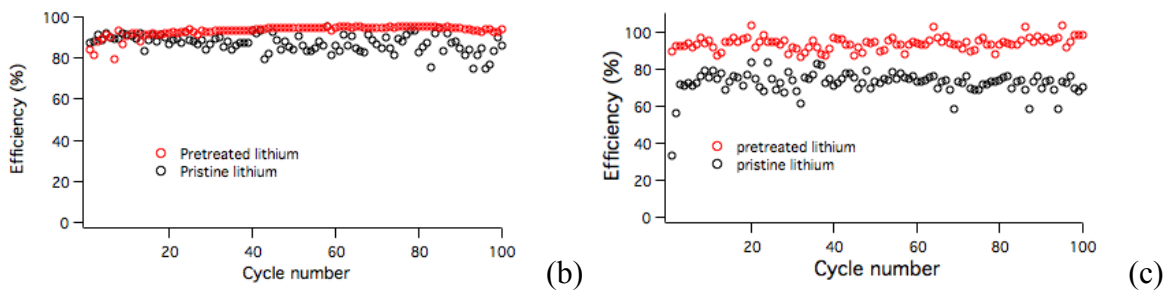

Figure 2. a) Short circuit time for symmetric cells at different current densities. The dark grey column represents the control case in which pristine lithium is used; the light grey column shows the short circuit time of the cell applying pretreated lithium metal. b) and c) Lithium deposition efficiency in Li/electrolyte/stainless steel cells comprised of pristine lithium and pretreated lithium metal, respectively. The cell is discharged for 30 min at a constant current density then recharged to 0.5V at the same current density. b) Current density = 0.2 mA/cm$^2$. c) Current density = 2 mA/cm$^2$.

The Coulombic efficiency (CE) provides a simple measure of the effectiveness of the Li surface protection offered by the AlI$_3$ treatment. CE was examined using a Li/electrolyte/stainless steel cell design, which allows the lithium loss on cycling to be accurately determined as the amount of lithium stripped divided by the amount of lithium plated on the stainless steel foil. A fixed amount of lithium is stripping from the lithium metal by a constant discharge current and deposited on to stainless steel, followed by a charge process where all the lithium is coming back to lithium electrode from stainless steel.[34] **Figure 2b and 2c** shows the Coulombic efficiency versus cycle number for cells with 1M LiTFSI in DOL/DME (v:v=1:1) at current densities of 0.2 mA/cm$^2$ and 2 mA/cm$^2$ (**Figure 2b and 2c** respectively). The black curve is the control case, in which pristine lithium is used, and the red curve represents the pretreated lithium. In **Figure 2b**, when pretreated lithium metal is used, the CE is stable at ~95%, while there is fluctuation in the cell with pristine lithium. There is still loss of efficiency with the protection of AlI$_3$ and that might be due to the exposed lithium on stainless steel reacting with the electrolyte. The improvement is still observed when the current density is increased to 2 mA/cm$^2$ (**Figure 2c**), where the efficiency is increased from ~70% to ~ 92% when lithium metal is pretreated with AlI$_3$. **Figure S5** reports the CE for electrolytes in which LiPS is directly added to promote parasitic reactions with the freshly deposited Li metal. The addition of LiPS to the electrolyte is observed to dramatically decrease the stability of the control cell, resulting in the large fluctuation of the efficiency when pristine lithium is used. However, consistent with what we found in **Figure 1**, the pretreated lithium metal shows improved stability against LiPS and the cycling is much more stable with an improved efficiency to ~91% over 100 cycles.

The results reported in **Figures 1 and 2** therefore demonstrate the effectiveness of AlI$_3$ as an additive in the formation of stable SEI layer on Li metal. The results also show that treatment of Li metal with AlI$_3$ greatly improves its stability against LiPS erosion, enhances its resistance to failure by lithium dendrite formation and improves the Coulombic efficiency of the cell. Before illustrating the benefits of these synergistic effects in a Li-S cell, we consider how AlI$_3$ performs these functions in detail.

**Figure S6** shows the impedance for symmetric cells before and after treatment with AlI$_3$. The solid lines correspond to the cells before cycling. The cells containing AlI$_3$ in the electrolyte already show lower impedance compared with the control case. We infer that there may already be a SEI layer formed before electrochemical treatment so that the conductive LiI layer leads to lower impedance. After a single cycle, the impedance of AlI$_3$ is observed to shrink substantially. This may be due to formation of a more complete SEI layer after the cleaning of lithium surface and the strong adsorption of I$^-$ anion on the electrodes. In contrast, for the control case, the impedance doesn't change too much before and after the treatment. It might be because after cleaning the lithium surface, it immediately it reacts with the electrolyte solvents and again forms lithium oxidation products, such as lithium hydroxide or other lithium oxides, which has similar interfacial chemistry and transport properties with the pristine surface and results in similar impedance.

The pretreated lithium metal was harvested, washed thoroughly with DOL/DME, and its surface chemistry characterized by means of XPS. **Figure 3a** shows a strong I 3d signal, which has well separated spin-orbit components with an energy separation of 11.5 eV. The deconvolution of I 3d 5/2 shows both Li-I and Al-I peaks, which is consistent with our hypothesis of strong I$^-$ ion adsorption. Also, Li 1s peak (**Figure 3b**) can be deconvoluted into Li-I, Li-Al alloy and Li-OH peaks.[35,36] We suspect that the hydroxides may be formed during the sample transfer or the reaction between lithium and the solvents. Again this is consistent with our hypothesis that there is formation of Li-I and Li-Al alloy during the discharge and charge process, which can help to stabilize lithium metal. Al signals are also detected in the XPS spectra. Deconvolution of Al 2s peak (**Figure 3c**) shows both the metal Al (120eV) [29] and Al ion (~118eV) peaks, [37] confirming the existence of Al metal and Al ion on the surface, providing clue for Li/Al alloy formation and Al$^{3+}$ adsorption on the electrode surface.

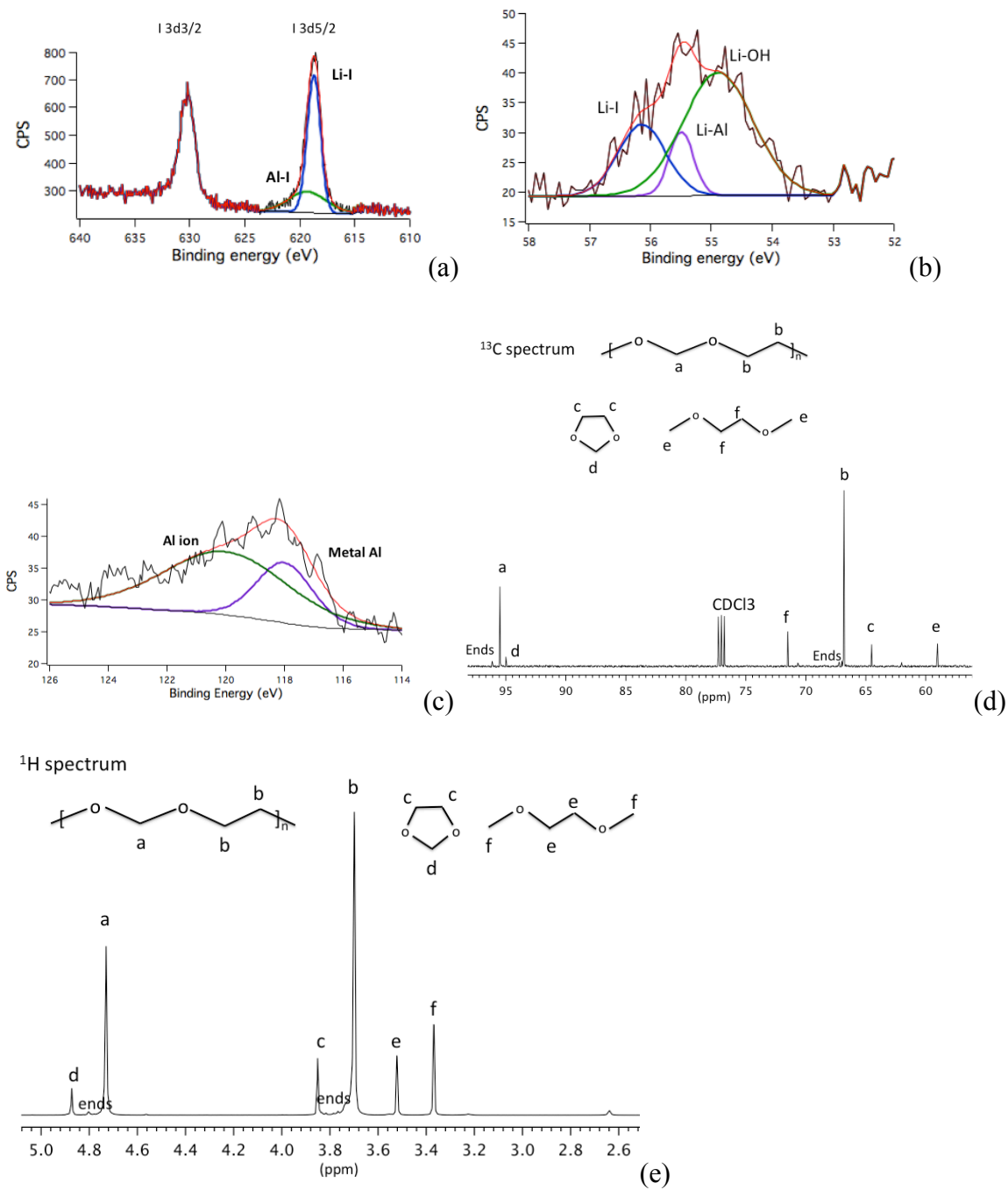

Figure 3. High resolution XPS analysis of the pretreated lithium metal. a) I 3d spectra; b) Li 1s spectra; c) Al 2s spectra. d) $^{13}$C NMR spectrum of the polymeric gel. e) $^1$H NMR spectrum of the polymeric gel.

Post-mortem characterization of the lithium metal was carried out via XPS and SEM, after galvanostatic polarization. It is seen that the XPS spectra of the pretreated material still shows the peaks for Al and I, which indicates that the SEI layer formed by Li and AlI$_3$ is very stable even after short circuit by polarization (**Figure S7b and S7c**). What's more interesting is that there is a film-like structure observed on the surface of the lithium metal (**Figure S8a**). After the short circuit happens in polarization test, the film structure seems to be punctured through by the lithium dendrite growth. The film was also observed to form spontaneously as an upper (lower-density phase) in an AlI$_3$ containing electrolyte (**Figure S8b**) after the electrolyte was rested for a period of about 2 weeks in an Ar-filled glove box. The gel-like membrane was separated from the liquid electrolyte and washed with DOL/DME to get rid of any interred salts and characterized by Gel Permeation Chromatography (GPC). The results reported in **Figure S9a** clearly show that the material is a polymer of molecular weight around 3200 with a polydispersity index of around 1.5.

In order to investigate the chemistry and structure of the polymer, mass spectra and NMR measurements were performed. **Figure S9b** shows that a series of species/fragments with a mass difference of 74 is observed. This mass increment is exactly the molecular weight of DOL, meaning that the film formed in the AlI$_3$-containing electrolytes is polyDOL. NMR analysis (**Figure 3d and 3e**) confirm that the gel is composed of polymerized DOL with a structure of n-[-O-CH2-O-CH2-CH2-]-n. Both the $^{13}$C and $^{1}$H spectra match pretty well with the proposed structure. They also show peaks for DOL and DME small molecules due to the solvent residue in the gel, which disappear when the gel is tested with diffusion ordered $^{1}$HNMR (**Figure S10**), where the polymer signal remain as slow diffusers. Thus the information provided by GPC, mass spectra and NMR spectra reveal that the gel is the product of DOL polymerization. Our finding while important is not surprising since the ring opening of DOL is already known to be initiated by Lewis acid acting as initiator and was one of the reasons we choose to work with $AlI_3$ as a strong Lewis acid able to attack the nucleophile center on the O atom and initiate the polymerization of DOL. [38-40]

**Figure 4** reports the electrochemical characteristics of Li-S cells based on the $AlI_3$-treated Li metal anodes at a current rate of 0.5C. The cathode used in these experiments was prepared by the methods reported earlier, where sulfur is infused into amine-functionalized CNT.[20] The green plot in **Figure 4a** corresponds to the control case, where pristine lithium metal is used, and the black curve represents pretreated lithium metal is used as anode. The capacity is very similar between the two cases while the efficiency is increased from 83%~92 %. What is more interesting is that when additional $AlI_3$ is introduced to the electrolyte, both the capacity and efficiency rise, with the CE exceeding 97% by the 100$^{th}$ cycle. We suspect that this benefit of utilizing additional $AlI_3$ in the Li-S electrolyte stems from the ability of $AlI_3$ to repair any defects in the surface coating formed during the pretreatment or by reaction with LiPS in the cycled electrolyte. The need for such repair is evident in **Figure 1d**, where the impedance of the symmetric cell using $AlI_3$-treated lithium also increases to some extent, suggesting the formation of a certain amount of $Li_2S$ on lithium surface. The voltage profile in **Figure 4b** shows that during cycling the overpotential in both the discharge and charge process is greatly suppressed when $AlI_3$ is incorporated. **Figure S11** shows the cyclic voltammetry of the Li-S battery applying the pretreated lithium as the anode. The peaks show typical characteristics of sulfur reduction and oxidization, and the stable position of the peaks indicates stable electrochemical reaction of sulfur.

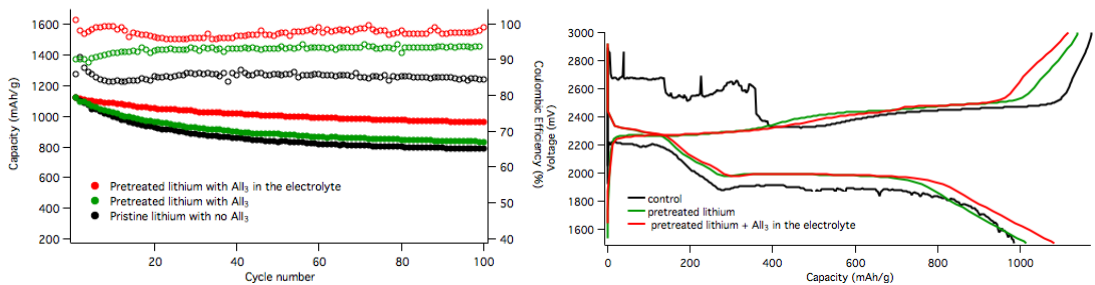

Figure 4. Electrochemical measurement of the Li-S battery using the pretreated lithium metal with $AlI_3$ as anode at current rate of 0.5C. a) Comparison of cycling performance of Li-S cells based on $AlI_3$-treated and untreated Li metal anodes. The black symbol represents the cell applying pristine lithium with additive free electrolyte; the green one represents the cell applying pretreated lithium with additive free electrolyte; the red one uses pretreated lithium in $AlI_3$ containing electrolyte. The blank symbol represents the corresponding Coulombic efficiency.  b) the corresponding voltage profile of the Li-S battery.

In conclusion, an electrochemical strategy involving $AlI_3$ as an additive is applied to form an artificial protection SEI layer on lithium metal, the component of which is verified to be LiI, Li-Al alloy and a thin polymer film due to DOL polymerization initialized by $Al^{3+}$. When the pretreated lithium metal is used as electrodes in symmetric cell or as anode in Li-S battery, the stability of the electrode is greatly improved by multiple synergistic processes, including the stability against LiPS, lithium dendrite resistance and the Coulombic efficiency of the cell. The promising electrochemical results and scientific understanding show the great promise of $AlI_3$ as an effective additive in the formation of a protecting SEI layer and also provide design principles and clues for other potential additives/strategies towards stable lithium anodes.

**Materials and Method**

**Pre-treatment of lithium metal with $AlI_3$.** A symmetric cell was assembled as Li/electrolyte/Li. 40μL 1M LiTFSI and 600ppm $AlI_3$ dissolved in DOL/DME (v:v=1:1)
18

was used as electrolyte. The symmetric cell was then discharged at a current of 2 mA/cm$^2$, at a depth of 2% DOD, followed by a charge process at the same current. The cell was opened in the glove box and the anode was used as protected lithium metal in other studies.

**Preparation of Li$_2$S$_8$ for LiPS stability study.** Li$_2$S$_8$ was prepared following Rauh[41] et.'s procedure in a solution process where stoichiometric amounts of elemental sulfur and Li$_2$S were co-dissolved into tetraglyme, followed by heating at 80$^\circ$C with stirring for 6h.

**Characterization:** The LiPS species in the electrolyte was detected by Shimadzu UV-Vis Spectrometer. Crystal structure was characterized using Scintag Theta-Theta X-ray Diffractiometer (XRD). Morphologies of the electrodes were studied using LEO 1550 FESEM (Keck SEM) and FEI Tecnai G2 T12 Spirit TEM (120 kV). Impedance was measured versus frequency using a Novocontrol N40 broadband dielectric spectroscopy. X-ray photoelectron spectroscopy (XPS) was used to do elemental analysis and obtain chemical bonding information. Waters Ambient-Temperature GPC was applied to do analysis of the molecular weight of polymer. The polymer was also investigated by the Direct Analysis in Real Time (DART) ambient ionization mass spectrometer. The structure of organic molecular was analyzed by INOVA 400 NMR facility.

**Electrochemical Characterization:** The sulfur cathode composite CNT-PEI/S is prepared as described in a previous study[20], and the sulfur content in the composite is 60%. 2030 coin-type cells were assembled using Lithium metal (0.76 mm. thick, Alfa Aesar) as the anode electrode, a microporous material, Celgard 2500, membranes as separator, a cathode with 80% as prepared CNT-PEI/S composite, 10% Super-P Li carbon black from TIMCAL, and 10% poly (vinylidene difluoride) (PVDF, Sigma Aldrich) as binder in an excess of N-methyl-2-pyrrolidone in NMP, and electrolyte of 40uL 1M lithium bis(trifluoromethanesulfone) imide (LiTFSI) for each cell. The sulfur loading per electrode is 1.2mg/cm$^2$. Cell assembly was carried out in an argon-filled glove-box (MBraun Labmaster). The room-temperature cycling characteristics of the cells wear evaluated under galcanostatic conditions using Neware CT-3008 battery testers and electrochemical processes in the celsls were studied by cyclic voltammetry using a CHI600D potentiostat.

## Acknowledgements

The authors acknowledge support of the National Science Foundation Partnerships for Innovation Program (Grant No. IIP-1237622). Electron microscopy, X-ray diffraction, X-ray photoelectron spectroscopy facilities and optical spectrometers available through the Cornell Center for Materials Research (CCMR) were used for this work (NSF Grant DMR-1120296).